\documentstyle[twoside,fleqn,npb,epsfig]{article}
\newcommand{\be}{\begin{equation}}
\newcommand{\ee}{\end{equation}}
\newcommand{\bea}{\begin{eqnarray}}
\newcommand{\eea}{\end{eqnarray}}
\newcommand{\beas}{\begin{eqnarray*}}
\newcommand{\eeas}{\end{eqnarray*}}
\newcommand{\bookfig}[5]{
\begin{figure}\centering\fbox{\epsfysize=#5cm \epsfbox{#1}}
\caption[#2]{#4}\label{#3}
\end{figure}
}

\newcommand{\AmS}{{\protect\the\textfont2
  A\kern-.1667em\lower.5ex\hbox{M}\kern-.125emS}}

\title{Feynman diagrams and polylogarithms: shuffles and pentagons}

\author{D.~Kreimer\address{Physics Department,
        Mainz University, \\
        D-55099 Mainz, Germany}%
        \thanks{Heisenberg Fellow, MZ-TH/00-21,
Talk given at {\em Loops and Legs in Quantum Field Theory},
Bastei, April 2000}}

\begin{document}

\begin{abstract}
We summarize the Hopf algebra structure on Feynman diagrams and
emphasize the interest in further algebraic structures hidden in
Feynman graphs.
\end{abstract}

\maketitle

\section{COMBINATORIAL RENORMALIZATION}
There is a universal combinatorial Hopf algebra structure hidden
in the process of renormalization \cite{1,2,3,5,4,6}. The universality
of this structure can be conveniently understood if one considers
the UV singularities of Feynman graphs from the viewpoint of
configuration spaces. Then, the combinatorial structure of
renormalization can be summarized as follows. Each Feynman diagram
has various sectors which suffer from short-distance
singularities. These sectors are stratified by rooted trees, from
which the Hopf algebra structures of \cite{1,2,3,5,4,6} are
obtained. Figure \ref{f1} gives an example.
\bookfig{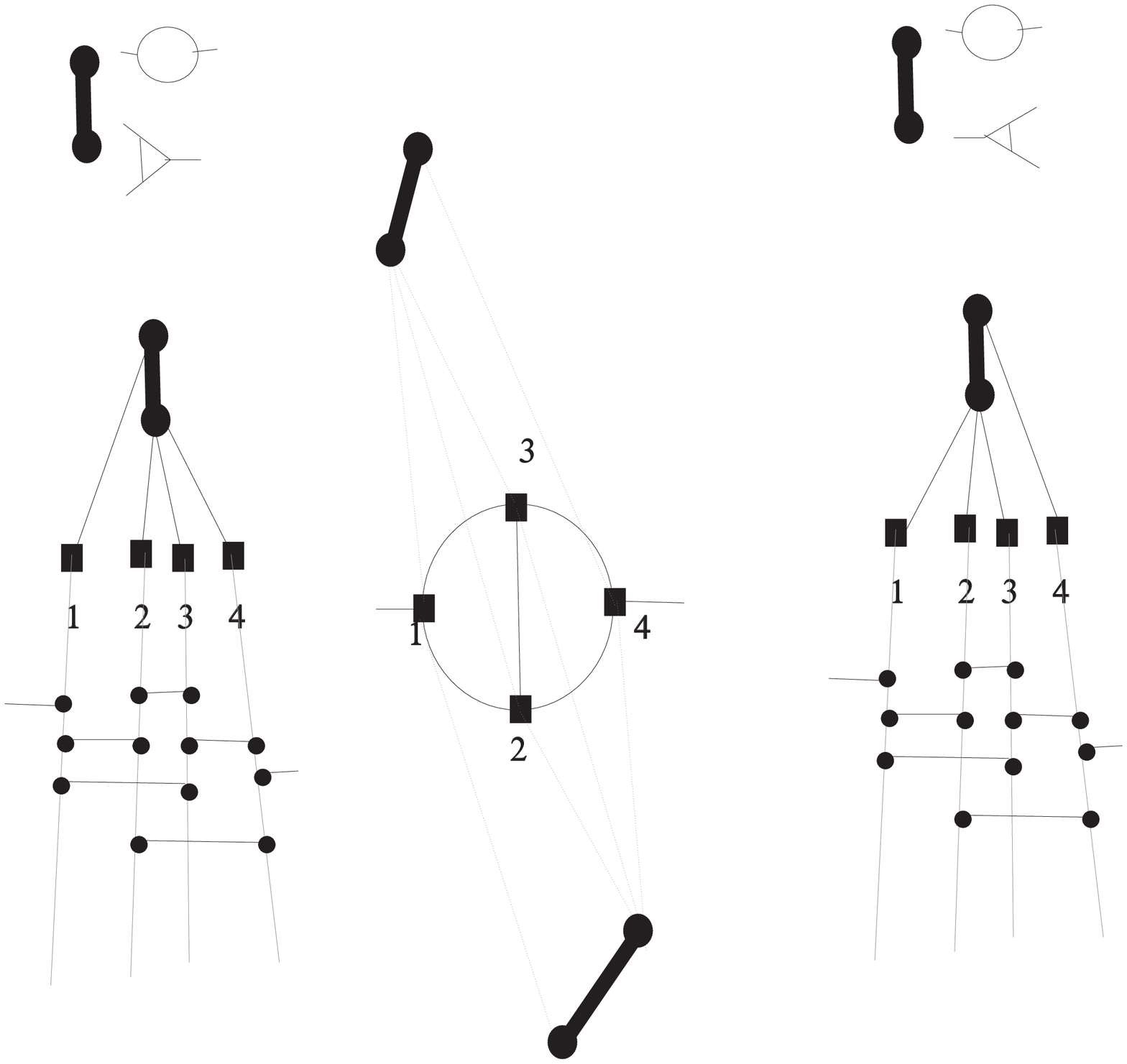}{}{f1}{A Feynman diagram with overlapping
divergences. Its short distance (UV) singularities are stratified
by two rooted trees as indicated. We can either describe the
divergent sectors by rooted trees decorated with primitive
diagrams, or in a manner closely related to the notations used by
Fulton and MacPherson \cite{FMac} in the study of configuration
spaces. In the latter case,
singular sectors appear when either the three vertices
$\{2,3,4\}$ approach a diagonal
first (with the final overall divergence
appearing when vertex $1$ approaches the same diagonal)
or when the vertices denoted $\{1,2,3\}$ come together first and then
approach
vertex $4$. In all cases, the rooted trees which govern the Hopf
algebra structure are the rooted trees with two vertices, here
given in bold black lines.}{6} The corresponding Hopf algebra can
be formulated on rooted trees or, equivalently, directly on
Feynman graphs. Details of the relation to configuration spaces
will be described elsewhere. Here, we want to use this
representation to motivate further investigation in algebraic
relations between Feynman graphs, and report some encouraging
first results which will be discussed in much greater detail in
future work.

Figure \ref{f1} essentially shows how the short distance
singularities are located in sectors stratified by rooted trees.
This is no surprise as the singularities are constrained to
(sub-)diagonals. Along such subdiagonals, one essentially
confronts the product of distributions with coinciding support.

This localization of singularities along diagonals
conveniently allows to rely on suitable local subtractions, whose
combinatorics can be described as Hopf algebra operations on the
trees of Figure \ref{f1}. It can also be directly described in
terms of Feynman graphs with coproducts of the form $$
\Delta(\Gamma)=\Gamma\otimes e + e \otimes \Gamma+
\sum_{\gamma{\subset\atop \not=}\Gamma}\gamma\otimes\Gamma/\gamma,$$ where the
relation to rooted trees can be read off from the figure.
Essentially, the sum over all rooted trees describing the various
divergent sectors in the graph replaces the graph when we go from
the Hopf algebra of graphs to the Hopf algebra of trees \cite{2}.

Here, $\Gamma/\gamma$ corresponds to a suitable collapse of
subgraphs to a point, which in the analytic
expressions corresponding to the graphs, furnishes an appropriate
insertions of local, polynomial, operators, a process which can be
conveniently formulated by considering distributions on the space
of external momenta of a graph \cite{6}.

Finally, the transition from an unrenormalized graph to a
renormalized graph can be summarized in a succinct formula. It
reads on graphs as the convolution $$
S_R\star\phi(\Gamma)=S_R(\Gamma)+\phi(\Gamma)+\sum_{\gamma
{\subset \atop
\not=}\Gamma} S_R(\gamma)\phi(\Gamma/\gamma),$$ where $R:V\to V$
specifies the renormalization scheme, and $\phi,S_R:H\to V$ are
algebra maps which assign to Hopf algebra elements, graphs,
analytic expressions in $V$ corresponding to the Feynman rules
($\phi$) and
to the counterterm ($S_R$).
$S_R$ appears here as a twist of the map $\phi\circ
S\equiv S_{R={\rm id}_V}: H\to V$ given by
$$S_R(\Gamma)=-R[\phi(\Gamma)+\sum_{\gamma{\subset \atop
\not=}\Gamma}
S_R(\gamma)\phi(\Gamma/\gamma)].$$

Thus, we obtain a group structure due to the fact that the counit,
coproduct and coinverse of the Hopf algebra provide a unit,
a product, and an
inverse for characters of the Hopf algebra.

This enables us to describe changes in renormalization schemes via an
obvious generalization of Chen's lemma \cite{3} $$
S_{R^\prime}\star\phi=S_{R^\prime}\star S_R\circ S\star S_R\star
\phi,$$ and similarly for a change of the character $\phi$, $$
S_{R}\star\phi^\prime=S_{R}\star \phi\star \phi\circ S\star
\phi^\prime.$$
 There is a distinguished renormalization scheme which further
enables
 us to identify the transition from the unrenormalized Green
 functions
 to the renormalized ones with the Birkhoff decomposition: the
 minimal subtraction scheme in dimensional regularization
 \cite{4,6}. This allowed Alain Connes and the author
 to gain further insight in the nature of
 renormalization: in its very essence, renormalization is about
 diffeomorphisms of physical observables. If one works this out
 for the
 simplest but generic
 instance of the diffeomorphism of the coupling constant,
 one confronts two Hopf algebra structures: the Hopf algebra
 delivered by the composition of diffeomorphisms \cite{CM}
 and the Hopf
 algebra structure of Feynman graphs, as we can regard the
 effective
 coupling constant as a series in graphs via $g_{\rm eff}(g)=gZ_1
 Z_3^{-3/2}$, and one ends up with the fact that this formula
 delivers a homomorphism of Hopf algebras.
Here, $Z_1\equiv Z_1(g)$, $Z_3\equiv Z_3(g)$ are both to be
regarded as invertible formal series in $g$ with
$Z_1(0)=Z_3(0)=1$.

 Upon the natural action of the rescaling group, this implies
 the 't Hooft conditions which ensure the existence of a
 well-defined $\beta$-function, which is a pull-back of the
 natural one-parameter group of rescalings in the above group
 \cite{6}.

These Hopf algebras can be made into efficient algorithms \cite{symb}
and to develop routines which possibly start
at the level of Wick contactions and completely automate
the BPHZ recursions for a given QFT remains as a desirable and
achievable challenge for computational physics. Along the same lines,
one should hope that in the long run the very fact that this Hopf
algebra
can be formulated in quite general set-theoretic terms will enable us
in the future to succinctly approach asymptotic expansions
and stratifications of regions \cite{smirn} in conceptually satisfying
and efficient manners, using appropriate conditions
to identify the subgraphs of interest for a particular asymptotic
expansion.

Having achieved in such a way a completely satisfactory
description of the multiplicative subtraction mechanism which
guarantees finiteness of renormalized Green functions, one is
tempted to ask for more structure which follows from the disclosed
Hopf algebra structure. This should lead us eventually
to a consideration of
the transcendent content of a field theory,
towards its polylogarithmic content which plays such a prominent
role in QFT as also this conference gives ample evidence.

\section{SHUFFLES}
\bookfig{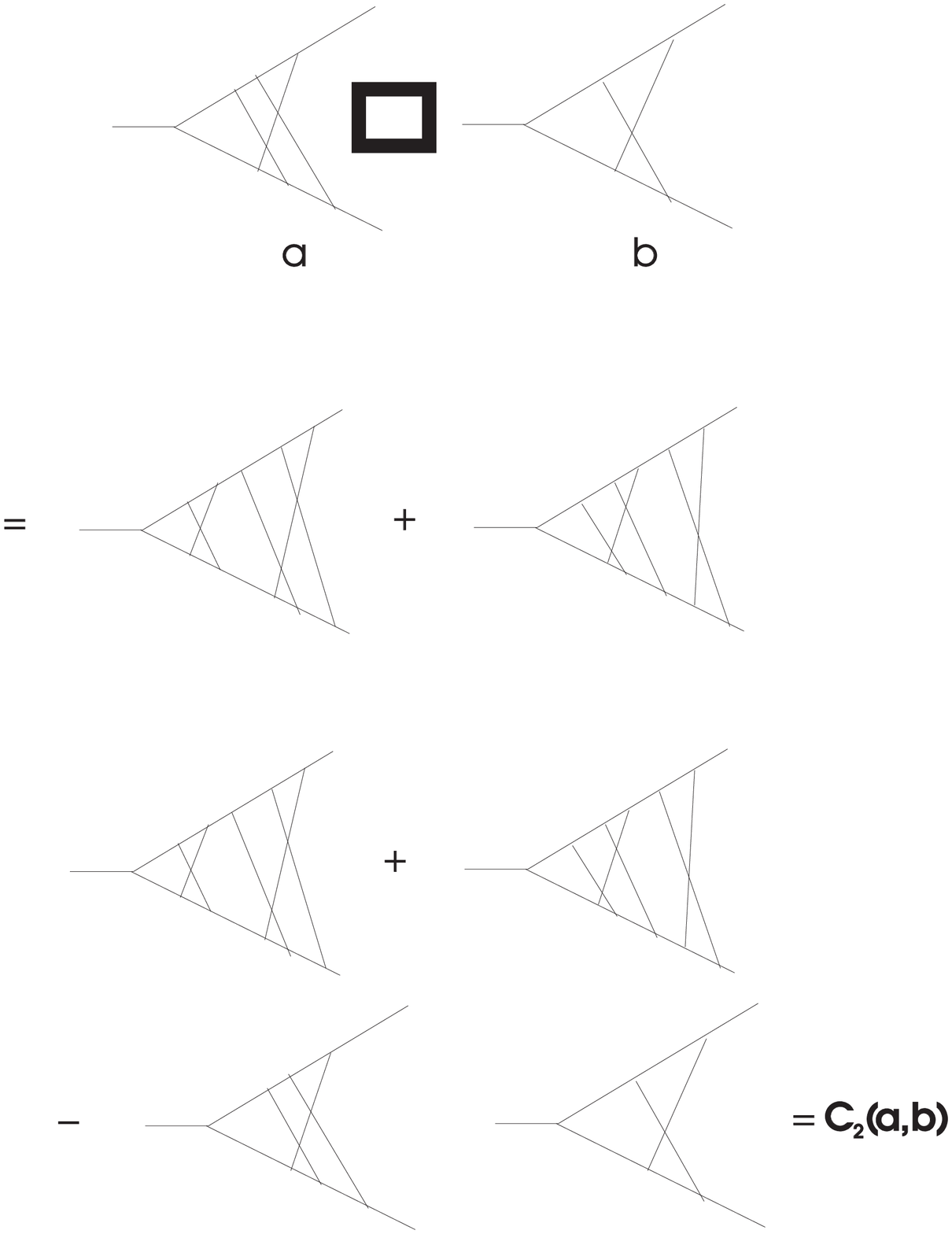}{}{f2}{Shuffling Feynman
graphs. In the first line,
we see the shuffle product of two primitive (subdivergence-free)
Feynman graphs denoted by $a,b$ (in $\phi^3$ theory in six
dimensions, say).
The shuffle delivers the sum of two diagrams given in the second line.
This shuffle fulfills a shuffle identity, which means that it is
distinguished
from the product of the two graphs only by a primitive contribution
$C_2(a,b)$, suggesting a quasi-shuffle identity.}{6}
 It is not difficult to see that the Hopf algebra structure of
Feynman diagrams allows to define shuffle algebras \cite{7}. To
this end, we can utilize the formulation in terms of rooted
trees, which allows to write the coproduct as $$ \Delta\circ
B_+^{(x)}(X)=B_+^{(x)}(X)\otimes e
+[{\rm id}\otimes B_+^{(x)}]\Delta(X).$$
Corresponding operators exist, due to a theorem in \cite{5}, in
the formulation in terms of Feynman graphs, and amount to plug the
graphs $X$ in the graph $x$ in all possible ways. One can even
consider operators $B_+^{x,i}(X)$ which correspond to plug in the
graph $X$ at the place $i$ in the primitive graph $x$. Here, a
place $i$ can be a specified internal line (if $X$ is a
self-energy graph for that type of line) or vertex (if $X$ is a
vertex correction graph) of $x$ which is replaced by $X$.

One can now define a (quasi-)shuffle product $\Box$
iteratively on graphs (for
fixed chosen places say) by defining \beas B_+^{(x)}(X){\bf
\Box}B_+^{(y)}(Y) & = & B_+^{(x)}(X{\bf \Box} B_+{(y)}(Y))\\ & &
+B_+^{(y)}(B_+^{(x)}(X){\bf \Box} Y)\\
 & & +B_+^{C_2(x,y)}(X{\bf \Box} Y
).\eeas If a shuffle identity $\phi(X\Box Y)=\phi(X)\phi(Y)$
holds up to $n$ loops, then the shuffle product is well-defined at
$n+1$ loops. Here, $C_2$ is a map which assigns to two primitive
Feynman graphs a new one
\cite{MEH}. Such shuffle algebras are provided by iterated integrals,
and hence in particular by polylogarithms as well as in the study of
MZVs and Euler sums.

To obtain a well-defined shuffle algebra, one needs commutativity
and associativity of this map. It is the latter which, in the case
of Yukawa theory, was explicitly constructed and shown to hold up
to finite parts \cite{7}. The finite violation of associativity
was only a function of the grading, the loop-numbers,
$n_x,n_y,n_z$, of the involved primitive graphs,
$$C_2(x,C_2(y,z))-C_2(C_2(x,y),z)$$ $$=(n_x-n_z)C(x,y,z)$$ with some
constant $C(x,y,z)=F(x)F(y)F(z)$ factorizing with respect
to the primitive graphs $x,y,z$.
This ensures the pentagon relation of Figure \ref{f3}.

 It is an open question if at the next order
in $(D-4)$ a pentagon relation still holds or if a higher coherence
law is needed. A full understanding of this question will provide
valuable insight as to what extent Feynman diagrams might
generalize the algebraic structures of the polylogarithm.
\bookfig{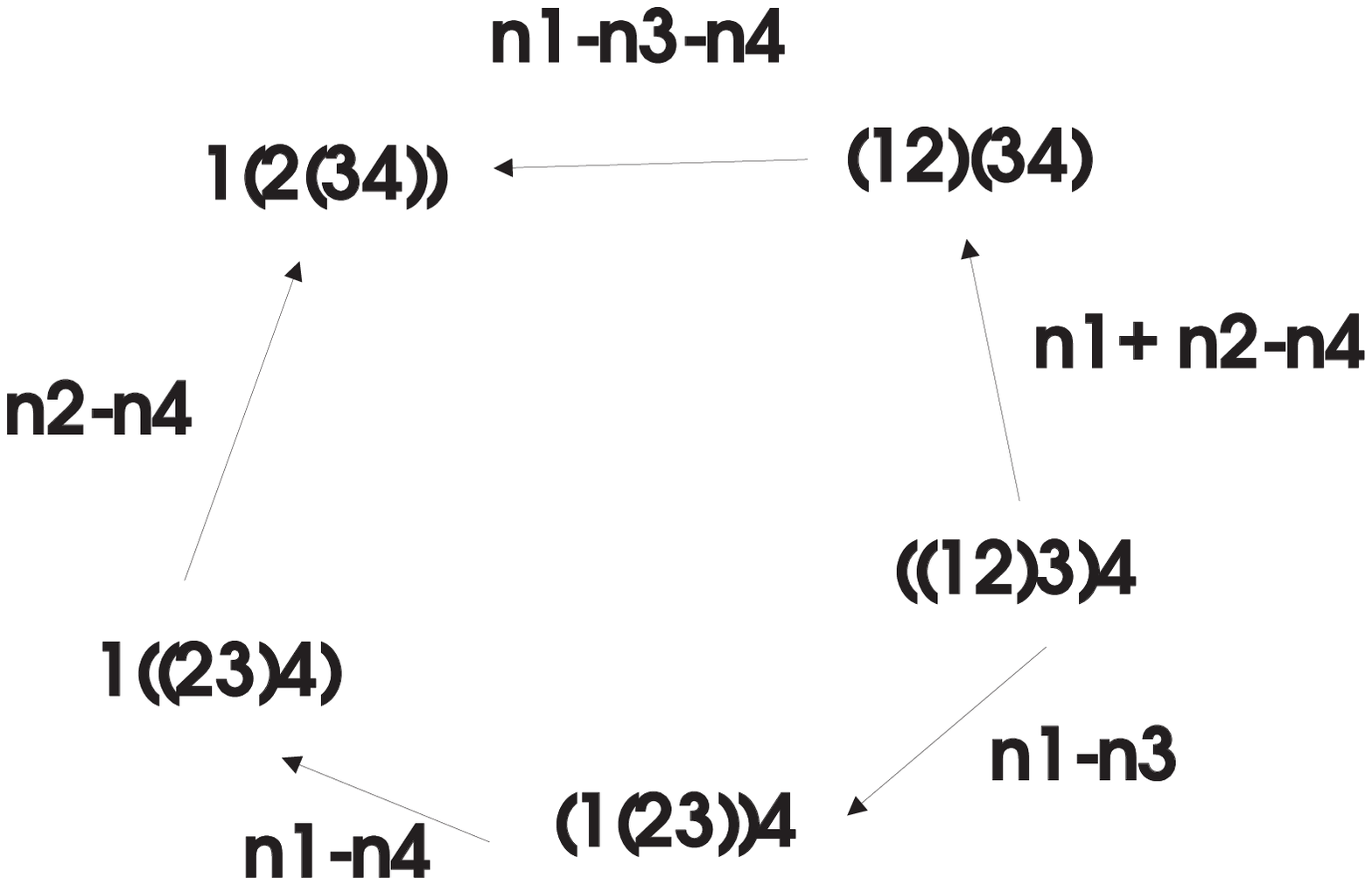}{}{f3}{The pentagon relation.
There are two ways to go from $C_2(C_2(C_2(x_1,x_2),x_3),x_4)
\equiv ((12)3)4$
to $C_2(x_1,C_2(x_2,C_2(x_3,x_4)))\equiv
1(2(34))$. Both should agree up to finite
parts. Here, a shift of a pair of brackets
is accompanied by the indicated difference in loop numbers
which add to the same total in both ways.}{4} Here is not the
place to go into details of the structure of polylogarithm, which
is a quite fascinating subject in its own right
\cite{poly}, also for a physicist \cite{old}.\footnote{The
webpage of Michael Hoffman collects some useful references to the
works of Zagier, Goncharov and many others.} The polylogarithm, as
an iterated integral, fulfills shuffle identities. But the
polylogarithm also relates to the Grothendieck Teichm\"uller
group, and the remarks in the next section can be regarded as the
first cautious steps to investigate similar structure in Feynman
graphs.
\section{MORE STRUCTURE}
Figure \ref{f1} by itself suggest to investigate more structure.
The Lie algebra structure which comes along with the Hopf algebra
of Feynman graphs is determined by a Lie bracket
$[\Gamma_1,\Gamma_2]$ which sums over all possible ways of
plugging a graph $\Gamma_1$ in $\Gamma_2$ and subtracts doing
it the other way round \cite{6}. Figure \ref{f4} gives an example.
\bookfig{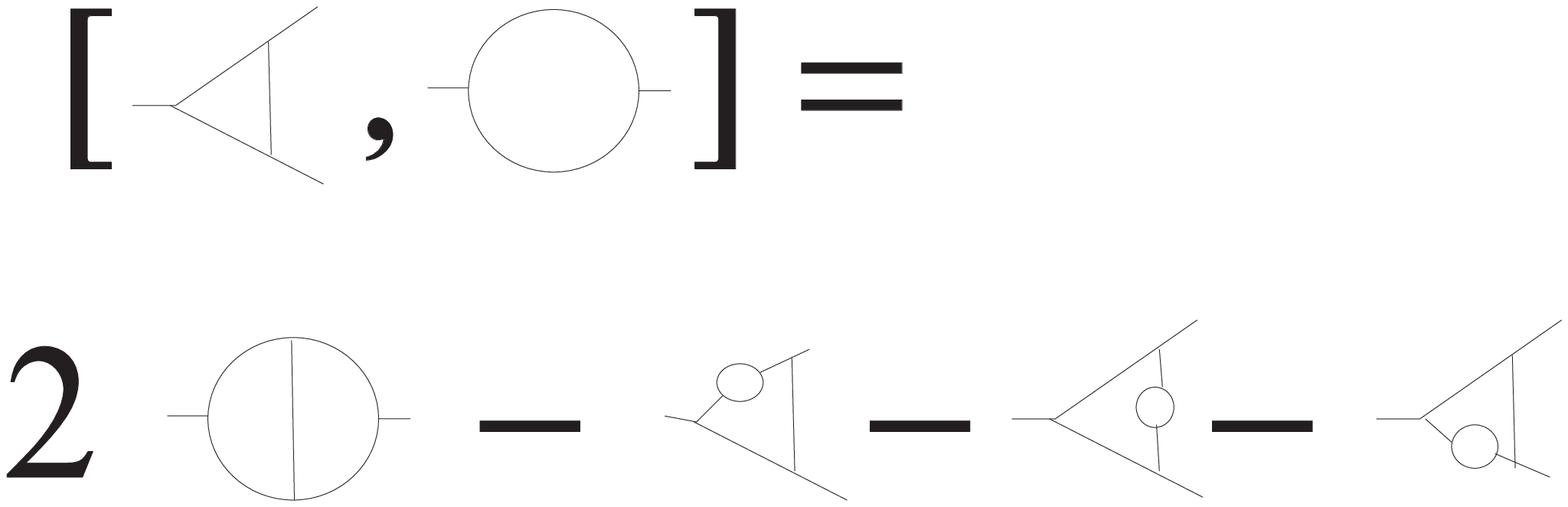}{}{f4}{The bracket of two graphs,
a vertex correction and a self-energy graph in this example.
The self-energy graph has two internal vertices. Both can be replaced
by the vertex correction, which delivers twice the first Feynman graph
in the second row. The other way round, any internal line of the
vertex
correction can be dressed by the self-energy graph, delivering the
remaining three graphs of the final row.}{2} There is a
more basic operation involved in this process, the insertion of a
graph $\Gamma_1$ at a chosen internal line or vertex of
$\Gamma_2$. In the Lie bracket, we involve only the sum over all
internal such places, and are thus completely insensitive to
structures which depend on these places. There is an obvious operad
structure involved when one labels internal lines and vertices,
and it is of interest to investigate Feynman graphs with respect
to this operad structure. Figure \ref{f5} gives an example
of plugging in a self-energy at a specified place.
\bookfig{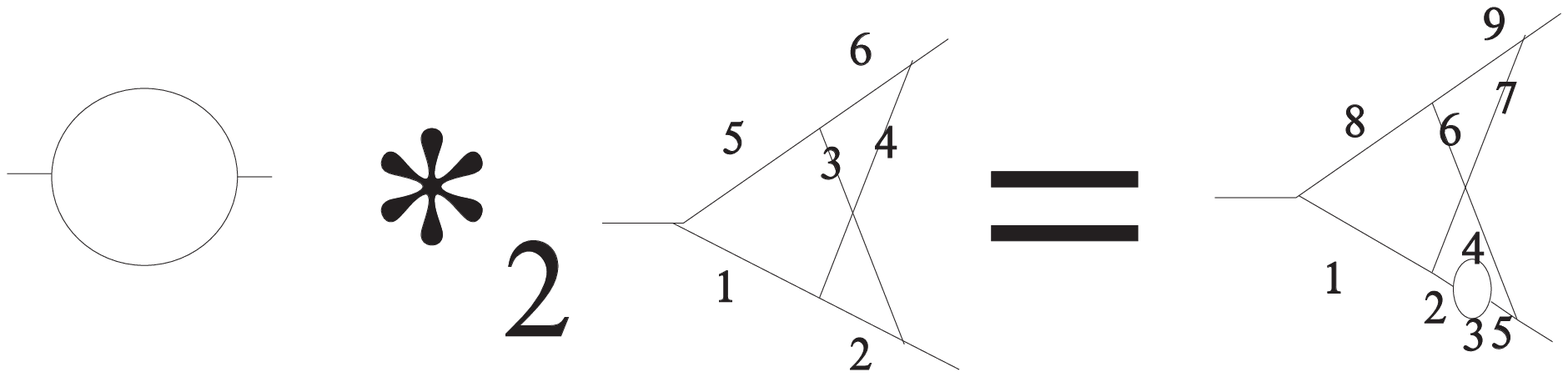}{}{f5}{Insertion of a self-energy at the specified
internal line 2 in a vertex correction.}{1.6}
\bookfig{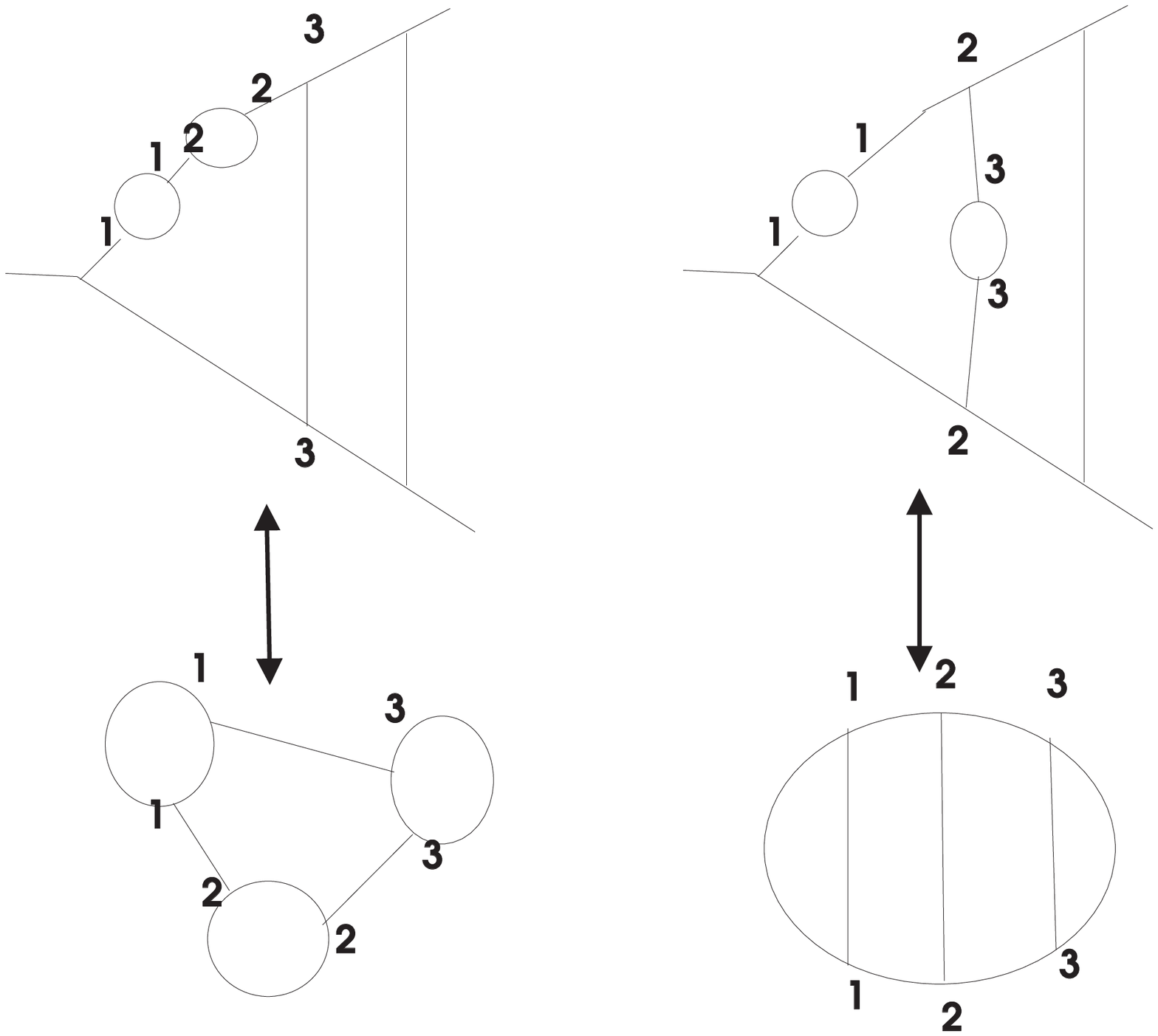}{}{f6}{These two graphs are only distinguished by the
  places
into which subdivergences are plugged, and their difference
has only a first order pole. They correspond to
different topologies, but have similar renormalization parts.
The difference in their topology is apparent when one reads them as
chord diagrams, providing different Gauss codes $\{1,1,2,2,3,3\}$
and the rational (ladder) code $\{1,2,3,3,2,1\}$ \cite{8}.
Upon evaluation, their difference is typically proportional to
$\zeta(3)/z$.}{4}
There is no room here for a detailed account
of such questions which has to be postponed to future work.
Suffices it to report on an interesting observation concerning
possibilities of plugging in a Feynman graph at two different
places in a graph. One of the simplest such instances can be
obtained by letting $\Gamma_2$ be a one-loop vertex function in
massless $\phi^3$-theory in six dimensions at zero momentum
transfer, and $\Gamma_1$ be some self-energy graph in this theory.
Then, there are two different places $a,b$ in $\Gamma_2$ into
which $\Gamma_1$ can be inserted, and the Hopf algebra element
$$p:=\Gamma_2[\star_a-\star_b]\Gamma_1$$ is primitive under the
coproduct:
$$
\Delta(p)=p\otimes e+e\otimes p,
$$
due to the fact that the coproduct structure (in the massless
theory) is insensitive to the place where a subgraph is located.
Hence, it can at most provide a first order pole
in $z$ (where $z$ is the complex deviation from the integer
dimension of spacetime in which the theory is to be evaluated).
Its evaluation reveals more: it is finite, it has
no negative part --no pole in $z$--
in its Birkhoff decomposition. More detailed
investigations suggest to investigate finiteness for primitive
elements of the form
$\Gamma_i[\star_a-\star_b]\Gamma_j[\star_{a^\prime}-\star_{b^\prime}]\Gamma_k$,
(where $a,b,a^\prime,b^\prime$ are places in $\Gamma_k$)
which will be reported in future work, fitting nicely with the
geometric picture
developped in \cite{6}.

As $p$ is finite, plugging a divergent subgraph into it
still cannot generate a negative part in the Birkhoff decomposition,
while plugging the so constructed graph as a subgraph in a primitive
graph delivers a new primitive element. Its evaluation
tests out the topologiocal differences between two graphs which have
the same substructure with regard to their renormalization parts,
as was already observed some time ago \cite{8}, with the most striking
observation being
that this measure of the difference in the topology is proportional
to $\zeta(3)$.

In \cite{6} Alain Connes
and the author presented a nice geometrical picture for the
Birkhoff decomposition of unrenormalized graphs. In light of the
kinship expressed in Figure \ref{f1} between singularities in
Feynman graphs and the study of generalized functions on
configuration spaces, algebraic relations like the above will
connect naturally to structures familiar from Grothendieck
Teichm\"uller groups, and we will investigate these connections in
the future in a hope to clarify  the role which the polylogarithm
plays in quantum field theory, in particular also with regard to gauge
theories.
\section*{Acknowledgements}
It is a pleasure to thank Johannes Bl\"umlein and
Tord Riemann for organizing this wonderful workshop,
and to thank all participants for the stimulating atmosphere.
As usual, I benefitted a lot from the tenacity and wisdom of my
collaborators David Broadhurst and Alain Connes.


\begin{thebibliography}{99}
\bibitem{1}
D.~Kreimer, Adv.\-Theor.\-Math.\-Phys.{\bf 2.2} (1998) 303;
q-alg/9707029.
\bibitem{2}
D.~Kreimer,
Comm.\-Math.\-Phys.{\bf 204} (1999) 669;
hep-th/9810022.
\bibitem{3}
D.~Kreimer,
Adv.\-Theor.\-Math.\-Phys.{\bf 3.3} (1999); hep-th/9901099.
\bibitem{5}
A.~Connes, D.~Kreimer, Comm.\-Math.\-Phys.{\bf 199}
(1998) 203; hep-th/9808042.
\bibitem{4}
A.~Connes, D.~Kreimer,
J.\-High Energy Phys.{\bf 09} (1999) 024,
hep-th/9909185.
\bibitem{6}
A.~Connes, D.~Kreimer,
Comm.\-Math.\-Phys.{\bf 210} (2000) 249;
hep-th/9912092;\\
A.~Connes, D.~Kreimer,
{\em Renormalization in quantum field theory and the Riemann--Hilbert
problem II: the $\beta$-function, diffeomorphisms and the
renormalization
group}, IHES/M/00-22; hep-th/0003188.
\bibitem{FMac}
W.~Fulton, R.~MacPherson, Ann.\-Math.(2) {\bf 139} (1994) 183.
\bibitem{CM}
A.~Connes, H.~Moscovici, Comm.\-Math.\-Phys.{\bf 198} (1998) 199.
\bibitem{symb}
D.J.~Broadhurst, D.~Kreimer, Phys.\-Lett.{\bf B475} (2000) 63;
hep-th/9912093;\\
D.J.~Broadhurst, D.~Kreimer, J.\-Symb.\-Comp.{\bf 27} (1999) 581;
hep-th/9810087.
\bibitem{smirn} V.A.~Smirnov, Phys.\-Lett.{\bf B465} (1999) 226;
hep-ph/9907471.
\bibitem{7}
D.~Kreimer, {\em Shuffling Quantum Field Theory},
to appear in Lett.\-Math.\-Phys.; hep-th/9912290.
\bibitem{MEH}
M.E.~Hoffman, J.\-Alg.\-Comb.{\bf 11} (2000) 49; math-qa/9907173.
\bibitem{poly}
 J.M.~Borwein and D.M.~Bradley and D.J.~Broadhurst,
Elec.~J.Comb.{\bf 4}(2), R5, (1997);\\
D.J.~Broadhurst, {\em Conjectured Enumeration of irreducible
Multiple Zeta Values, from Knots and Feynman Diagrams},
Phys.Lett.{\bf B}, in press, hep-th/9612012;\\ D.J.~Broadhurst,
{\em On the enumeration of irreducible $k$-fold Euler sums and
their roles in knot theory and field theory}, J.Math.Phys., in
press, hep-th/9604128;\\ A.~Goncharov, {\em Multiple
Zeta Values, Galois groups, and geometry of modular
varieties}, math.ag/0005069;\\
A.~Goncharov, Math.\- Res.\- Lett.{\bf 4}, (1997) 617;\\
R.M.~Hain, {\em Classical polylogarithms}, American Mathemtical
Society, Proc.\-Symp.\-Pure Math.{\bf 55}, Pt. 2, (1994) 3;\\
M.E.~Hoffman, J.\-Alg.{\bf 194}, (1997) 477;\\
D. Zagier, in  First European
Congress of Mathematics, Vol. II, Birkhauser, Boston, 1994,
497-512.
\bibitem{old}
D.J.~Broadhurst, D.~Kreimer, Int.\-J.\-Mod.\-Phys.{\bf C6}  (1995)
519;\\
D.J.~Broadhurst, D.~Kreimer, Phys.\-Lett.{\bf B393}
(1997) 403; hep-th/9609128.
\bibitem{8}
D.~Kreimer, Eur.\-Phys.\-J.{\bf C2} (1998) 757; hep-th/9610128.
\end{thebibliography}
\end{document}